%% file: main.tex
\documentclass[sigconf]{acmart}
\usepackage{booktabs} % For formal tables
\usepackage{tablefootnote}
\copyrightyear{2018} 
\acmYear{2018} 
\setcopyright{acmcopyright}
\acmConference[SIGIR'18]{41st International ACM SIGIR Conference on Research and Development in Information Retrieval}{July 8--12, 2018}{Ann Arbor, MI, USA}
\acmBooktitle{SIGIR '18: 41st International ACM SIGIR Conference on Research and Development in Information Retrieval, July 8-12, 2018, Ann Arbor, MI, USA}
\acmPrice{15.00}
\acmDOI{10.1145/3209978.3210142}
\acmISBN{978-1-4503-5657-2/18/07} 
\title{Sanity Check: A Strong Alignment and Information Retrieval Baseline for Question Answering}

\author{Vikas Yadav}
\affiliation{%
  \institution{School of Information \\ University of Arizona}
}
\email{vikasy@email.arizona.edu}

\author{Rebecca Sharp}
\affiliation{%
  \institution{Department of Computer Science\\ University of Arizona}
}
\email{bsharp@email.arizona.edu}

\author{Mihai Surdeanu}
\affiliation{%
  \institution{Department of Computer Science\\ University of Arizona}
}
\email{msurdeanu@email.arizona.edu}

\begin{document}

\begin{abstract}
While increasingly complex approaches to question answering (QA) have been proposed, the true gain of these systems, particularly with respect to their expensive training requirements, can be inflated when they are not compared to adequate baselines. 
Here we propose an unsupervised, simple, and fast alignment and information retrieval baseline that incorporates two novel contributions: a \textit{one-to-many alignment} between query and document terms and \textit{negative alignment} as a proxy for discriminative information.  Our approach not only outperforms all conventional baselines as well as many supervised recurrent neural networks, but also approaches the state of the art for supervised systems on three QA datasets. 
With only three hyperparameters, we achieve 47\% P@1 on an 8th grade Science QA dataset,  32.9\% P@1 on a Yahoo! answers QA dataset and 64\% MAP on WikiQA. We also achieve 26.56\% and 58.36\% on ARC challenge and easy dataset respectively. In addition to including the additional ARC results in this version of the paper, for the ARC easy set only we also experimented with one additional parameter -- number of justifications retrieved.

\end{abstract}

%
% The code below should be generated by the tool at
% http://dl.acm.org/ccs.cfm
% Please copy and paste the code instead of the example below.
%



\maketitle

\input{body}

\bibliographystyle{ACM-Reference-Format}
\bibliography{refs}

\end{document}

%% file: body.tex
\section{Introduction}
\label{sec:intro}

Question answering (QA), i.e., finding short answers to natural language questions, is a challenging task that is an important step towards natural language understanding \cite{Etzioni:11}.  
With the recent and widespread success of deep architectures in natural language processing (NLP) tasks \cite{young2017recent}, more and more QA tasks have been approached with deep learning and in many cases the state of the art for a given question set is held by a neural architecture (e.g., \citet{Tymoshenko2017RankingKF} for WikiQA \cite{yang2015wikiqa}). %\todo{I don't think Tymoshenko uses a NN arch, no?}).
%becky: yes, with tree kernals too
However, with these architectures becoming the expectation, comparisons to \textit{strong} baselines often are neglected and thus allow us to lose sight of the true gain of these complex architectures, especially relative to their steep training costs. 

Here we introduce a strong alignment and information retrieval (IR) baseline that is simple, completely unsupervised, and trivially tuned. 
%Our approach approaches state-of-the-art supervised performance on three separate QA tasks, outperforming several more complex, supervised systems. 
%This finding strongly suggests that more context is needed for properly evaluating proposed architectures.
%
Specifically, the contributions of this work are:
{\flushleft {\bf (1)}}
{
We propose an unsupervised alignment and IR approach that features one-to-many alignments to better control for context, as well as negative alignments as a proxy for discriminative information.  We show that depending on the statistics of the given question set, different ratios of these components provide the best performance, but that this tuning can be accomplished with only three hyperparameters.}

{\flushleft {\bf (2)}} We demonstrate that our approach yields near state-of-the-art performance on four separate QA tasks, outperforming all baselines, and, more importantly, several more complex, supervised systems. 
These results suggest that, contrary to recent literature, unsupervised approaches that rely on simple bag-of-word strategies remain powerful contenders on QA tasks, and, minimally, should inform stronger QA baselines. The code to reproduce the results in this paper is publicly available\footnote{{\url{https://github.com/clulab/releases/tree/master/sigir2018-sanitycheck}}}. 
%Baselines used to evaluate performance gains of complex QA systems can be unnecessarily weak.  Here, we propose an alternative baseline that significantly outperforms all previous baselines and nearly reaches the state of the art for supervised systems trained on the same data.  using this info, closing gap between baselines and supervised systems... more fair and realistic a analysis
%Negative information - Our proposed model uses negative information in addition to positive information for choosing answers. We added the negative information to the positive information because relevant answers will have less negative information compared to other non-relevant questions for a question 

%\item Similarity to simple IR systems like BM25 - We use IDF, term frequency and word embeddings of words in query and candidate answers to calculate the ranking score. Since word embeddings also capture semantic information of a word, our proposed model outperforms all the previously defined baselines for QA on the three dataset.

\section{Related Work}
\label{sec:relatedwork}

%\todo{cite the orig, plus the sliding window one plus others that use same idea}
Information retrieval (IR) systems \cite[e.g.,][]{BM25} have served as the standard baseline for QA tasks \cite[][\textit{inter alia}]{Mihai_IR_QA,surdeanu:11}.  
%Conventional IR systems like BM25  have been used as baselines \cite{sharp2017tell} in various QA datasets like Kaggle 8th grade science QA. 
However, the lack of lexical overlap in many QA datasets %demonstrate a lack of lexical overlap 
between questions and answers \cite{Berger:00,fried2015higher,Zhou2015LearningCW}, makes standard IR approaches that rely on strict lexical matching less applicable.  %Addressing this limitation, 
 Several IR systems have been modified to use distributional similarity to align query terms to the most similar document term for various tasks, including document matching \cite{bridging_the_gap}, short text similarity \cite{kenter2015short}, and answer selection \mbox{\cite{chakravarti2017improved}}.   
However, using only a \textit{single} most similar term can lead to spurious matches, e.g., with different word senses.  Here we expand on this by allowing a \textit{one-to-many} mapping between a question term and similar answer terms to better represent how on-context a given answer candidate is.  

Negative information has also been show to be useful in answer sentence selection \cite{yih13,wang2016sentence}.
%\citet{LCLR} proposed an answer sentence selection approach using lexical semantic resources covering synonyms, antonyms, and hypernyms of words. \citet{wang2016sentence} employ a convolutional neural network (NN) to extract features from negative as well as positive lexical semantic information and reported a significant improvement with the inclusion of the negative information. 
%In our unsupervised framework, 
We also include negative information in the form of {\em negative alignments} to aid in distinguishing correct answers from close competitors.
% \cite{fried2015higher}.
% Here we include negative alignments as approximations of discriminative information to .
%Recent development in word embeddings using unsupervised models have opened opportunities for utilizing semantic information of words. Recently many researchers have used word embeddings for various tasks like document matching \cite{bridging_the_gap}, short text or sentence similarity \cite{kenter2015short}, 
%Few tasks like answer selection \cite{chakravarti2017improved} have used the same system \cite{bridging_the_gap} for answer selection.

{Several QA approaches have used similar features for establishing strong baseline systems, \cite[e.g.,][]{surdeanu:11,molino2016social}.  These systems are conceptually related to our work, but they are supervised and employed on different tasks, so their results are not directly comparable to our unsupervised system.}
%say look at similar features, but these models are supervised.  They are conceptually related but supervised and over different tasks, so results are not directly comparable.  (maybe a footnote)

\begin{figure}[t]
\begin{center}
\includegraphics[width=0.4\textwidth]{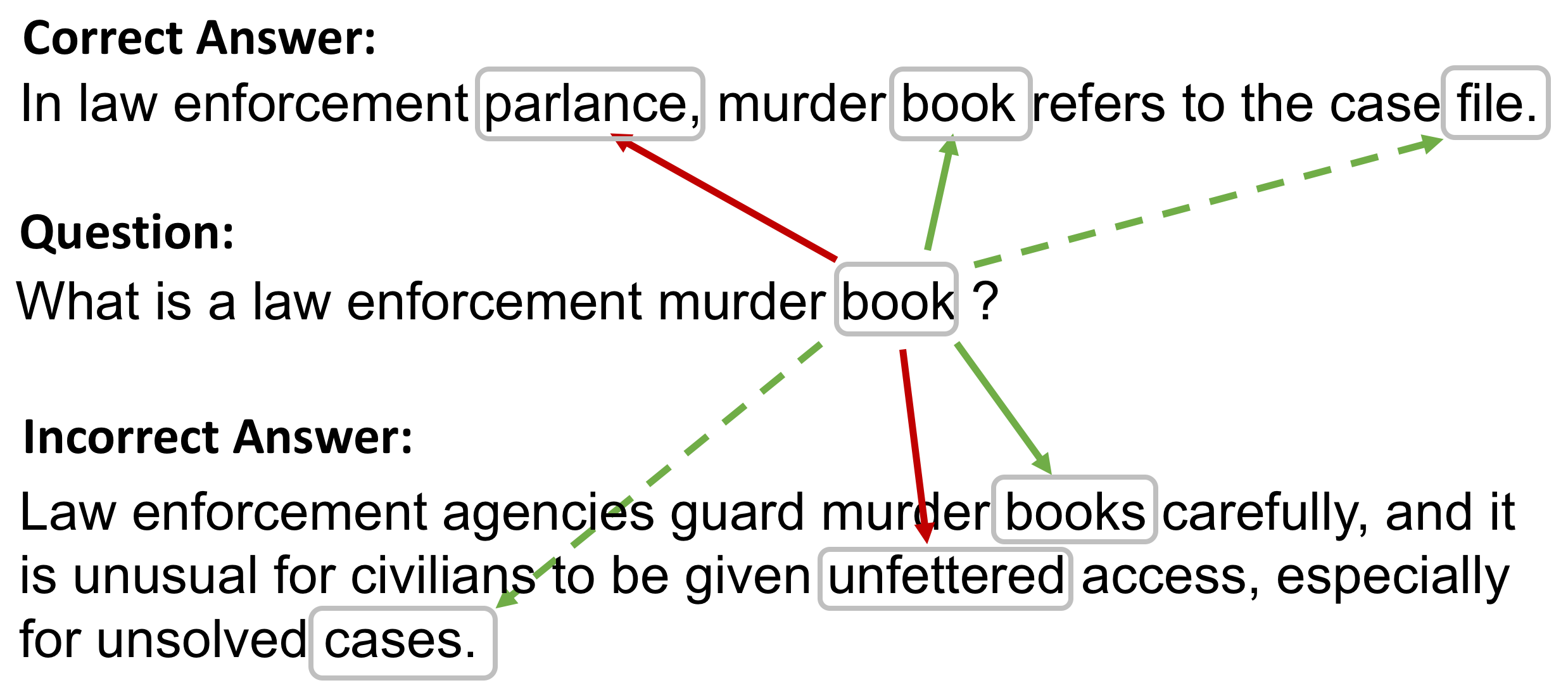}
\vspace{-4mm}
\caption{\footnotesize Example of our alignment approach for mapping terms in the question to the most similar and different terms in the candidate answer. 
 Highest-ranked alignments are shown with a solid green arrow, second-highest with a dashed green arrow, and lowest-ranked alignments are shown with a red arrow.}
\vspace{-4mm}
\label{fig:example}
\end{center}
\end{figure}

\section{Approach}
\label{sec:approach}
% \subsection{Preprocessing}
Our unsupervised QA approach is designed to robustly estimate how relevant a candidate answer is to a given question.  We do this by utilizing both positive and negative one-to-many alignments to approximate context.
Specifically, our approach operates in three main steps: 
% \todo{example, with real/artificial example, show how 1-many works and negative works. }\vikas{I have done it, I just need to paste it here.}

{\flushleft \textbf{(1) Preprocessing:}} We first pre-process both the question and its candidate answers using NLTK \cite{Bird_NLTK} and retain only non-stopword lemmas for each.  We additionally calculate the inverse document frequency (idf) 
%\cite[e.g.][Ch. 6]{manning08} 
of each query term, $q_i$ locally using:
\begin{small}
\begin{equation}
\mathit{idf}(q_i) = \log \frac{N-\mathit{docfreq}(q_i)+0.5}{\mathit{docfreq}(q_i) + 0.5}
\end{equation}
\end{small}
where $N$ is the number of questions and $docfreq(q_i)$ is the number of questions that contain $q_i$.
%\vspace{-5mm}

%After tokenization using NLTK \cite{Bird_NLTK} word tokenizer, we convert each token in lowercase. We then lemmatize each word using NLTK lemmatizer and then we remove all the stop words from the sentence using NLTK english stop words list.    
  
%  \item Next, for each candidate answer, we calculate the similarity between each term in the query and the answer candidate.  For this we use cosine similarity with of-the-shelf 300-dimensional Glove word embeddings \cite{pennington2014glove}.  This produces a matrix of alignments, $\mathcal{A} \in \mathbb{R}^{n,m}$, where $n$ and $m$ are the lengths of the question and candidate answer, respectively.
  
{\flushleft \textbf{(2) Alignment:}} Next we perform the one-to-many alignment between the terms in the question, $Q$, and the candidate answer, $A$. 
For $q_i \in Q$, we rank the terms $a_j \in A$ by their similarity to $q_i$ as determined by cosine similarity using off-the-shelf 300-dim Glove word embeddings \cite{pennington2014glove}, {which were not trained on any of the datasets used here.} For each $q_i$ we find the ranked top $K^+$ most similar terms in $A$, $\{ a^+_{q_i, 1}, a^+_{q_i, 2}, ..., a^+_{q_i, K^+} \}$ as well as the $K^-$ least similar terms, $\{ a^-_{q_i, 1}, a^-_{q_i, 2}, ..., a^-_{q_i, K^-} \}$.  
For example, in Figure \ref{fig:example}, \textit{book} in the question is aligned with \textit{book} and \textit{files} in the correct answer and with \textit{book} and \textit{case} (after preprocessing) in the incorrect answer as positive alignments.  
% \vspace{-1mm}

%We do not use the words which do not appear in Glove vectors vocabulary.
%bs: removed for space
  
%  \item We have used Glove word embeddings \cite{pennington2014glove} of 300 dimensions for each word in all the 3 datasets. We do not use the words which do not appear in Glove vectors vocabulary. 

%\item We have calculated the IDF’s of each term using the formula
%\todo{Becky, please add the IDF formula here }

%And we have calculated IDF locally within the dataset and hence no external knowledge is used throughout the experiments. We have experimented with both IDF’s of query terms and IDF’s of answer terms. Multiplying IDF of both query and answer term had lowered the performance in each experiment since IDF contribution was considered twice.

{\flushleft \textbf{(3) Candidate Answer Scoring:}} We then use these alignments along with the idfs of the question terms to find the score for each candidate answer, $s(Q,A)$, based on the weighted sums of the individual term alignment scores, such that:
%\vspace{-4mm}
\begin{small}
\begin{align}
s(Q, A) = \sum_{i=1}^N \mathit{idf}(q_i) \cdot \mathit{align}(q_i, A) \label{eqn:score}\\
\mathit{align}(q_i, A) = \mathit{pos}(q_i, A) + \lambda \cdot \mathit{neg}(q_i, A) \label{eqn:align} \\
\mathit{pos}(q_i, A) = \sum_{k=1}^{K^+} \frac{1}{k} \cdot a^+_{q_i, k} \\
\mathit{neg}(q_i, A) = \sum_{k=1}^{K^-} \frac{1}{k} \cdot a^-_{q_i, k} 
\end{align}%\right.
\end{small}
where $N$ is the number of question terms, $align(q_i, A)$ is the alignment score between the question term, $q_i$ and the answer candidate, $A$, and $\lambda$ is the weight for the negative information.  $pos(q_i, A)$ and $neg(q_i, A)$ represent the scores for the one-to-many alignments for the most and least similar terms respectively.  Importantly, the only hyperparameters involved are: $K^+$, $K^-$, and $\lambda$.

The intuition behind this formula is that by aggregating several alignments (i.e., through summing), the model can approximate context.  In terms of the example in Figure \ref{fig:example}, the secondary alignments for \textit{book} help discern that the correct answer is more on-context than the incorrect answer (i.e., \textit{book} is more similar to \textit{file} than it is to \textit{cases}).  Further, the negative alignments cause candidate answers with more off-context terms to be penalized more (as with \textit{book} and \textit{unfettered}).  These negative alignment penalties serve as an inexpensive proxy for discriminative learning.
%\todo{Explain in plain English why one-to-many and negatives are important.}

\section{Experiments}
\label{sec:experiments}

\subsection{Data}
We evaluate our approach on three distinct datasets:\footnote{As our approach is unsupervised, we tuned our two hyperparameters on the training and development partitions of each dataset.}

{\flushleft{\bf WikiQA:\footnote{\url{https://www.microsoft.com/en-us/download/details.aspx?id=52419}}}} a dataset created by \citet{yang2015wikiqa} for open-domain QA consisting of Bing queries and corresponding answer sentences taken from Wikipedia articles.  
The set is divided into train/dev/test partitions with 1040, 140 and 293 questions respectively.

{\flushleft{\bf Yahoo! Answers\footnote{\url{http://answers.yahoo.com}} (YA):}} 10,000 \textit{How} questions, each with a community-chosen best answer.\footnote{The questions are filtered to have at least 4 candidate answers, with an average of 9.}  We use the same 50-25-25 train/dev/test partitions as \citet{jansen14}. 

{\flushleft{\bf 8th Grade Science (ScienceQA):}}  
a set of multiple-choice science exam questions, each with four candidate answers.  We use the same 2500/800 train/test split as \cite{sharp2017tell}.  For better comparison with previous work, here we modify the approach slightly to score candidate answers against the same external knowledge base (KB) of short flash-card style texts from StudyStack\footnote{\url{https://www.studystack.com/}} and Quizlet\footnote{\url{ https://quizlet.com/}} as was used by \citet{sharp2017tell}.
Specifically, we first build IR queries from the question combined with each of the multiple-choice answers, and use there queries to retrieve the top five documents from the KB for each answer candidate.  
We then score each of these documents, as described in Section \ref{sec:approach}, using the combined question and answer candidate in place of $Q$ and each of the five documents in place of $A$.
The score for the answer candidate is then the sum of these five document scores.
%We then score each of these documents, $D$, as in Equation \ref{eqn:score}, modified such that $s(Q,A)$ becomes $s(Q+A, D)$.  The score for the answer candidate is then the sum of these five document scores.
%and score each of the top 50 documents from our retrieval step.  We then choose the candidate answer with the highest retrieval score.
%\todo{Becky, I am confused here. We have top 50 justifications from BM25 and we then use top 5 justifications to run our approach for each Cand.answer. We take sum of cosine similarity score for each justification, hence we add 5 similarity scores for each CA}

% \becky{Vikas -- is this what you did? please check.}  CORRECT
%\becky{Mihai/Vikas -- any chance we can move this mess to a footnote?}
% ms: no, this is important

{\flushleft{\bf ARC dataset:} \citet{clark2018think} presented the multiple choice AI2 Reasoning Challenge (ARC) dataset which is divided into two partitions: a Challenge set and an Easy set.  Each set is further divided into train, development, and test folds: \{1119, 299, 1172\} questions in train, development, and test folds respectively in the Challenge set and \{2251, 570, 2376\} questions in the Easy set. The KB for this dataset is provided by \citet{clark2018think} and covers approximately 95\% of the questions. As with ScienceQA, here we use the modified approach that uses five documents retrieved from the accompanying KB.
\subsection{Baselines}
 
We compare against the following baselines:

{\flushleft{\bf BM25:}} We choose the candidate answer with the highest BM25 score \cite{BM25}, using the default values for the hyperparameters.  

{\flushleft{\bf IDF Weighted Word Count:}}
We also compare against baselines from previous work based on \textit{tf-idf}. For WikiQA this is the IDF weighted word count baseline of \citet{yang2015wikiqa} and in YA this is the CR baseline of \citet{jansen14}.  In YA we also compare against the stronger supervised CR + LS baseline of \citet{jansen14}, which combines \textit{tf-idf} features with lexical semantic features into a linear SVM.
%. We have compared our model with weighted word count \cite{yang2015wikiqa} (which approximates as BM25 model) for WikiQA where IDF of question terms are used to weight the count of query terms appearing in candidate answer text. 
%CR \cite{jansen14} model which computes cosine similarity weighted with TF and IDF between answer candidates and question lemmas is taken as baseline for YA. We have also compared CR + LS features model \cite{jansen14} which is a stronger baseline for YA. 

{%\flushleft{\bf Candidate Retrieval (CR) and Lexical Semantics (LS)}}CR \cite{jansen14} model which computes cosine similarity weighted with TF and IDF between answer candidates and question lemmas is taken as baseline for YA. We have also compared CR + LS features model \cite{jansen14} which is a stronger baseline for YA.

{\flushleft{\bf Learning constrained latent representations (LCLR):}} For WikiQA, we also compare against LCLR \cite{yih13}, which was used as a strong baseline by \citet{yang2015wikiqa} to accompany the WikiQA dataset. LCLR uses rich lexical semantic information including synonyms, antonyms, hypernyms, and a vector space model for semantic word similarity.%, polarity-inducing latent semantic analysis (PILSA)\cite{yih2012polarity} for synonyms etc. 

{\flushleft{\bf AI2 IR Solver:}} The IR solver of \citet{clark2016combining} ranks according to the IR score where each retrieved document must contain at least one non-stop word from each of the question and the candidate answer.

{\flushleft{\bf Single-Alignment (One-to-one):}} 
We use only the single highest scoring pair of query word and answer word, i.e., $K^+ = 1$ and $K^- = 0$. 
%We compute the similarity scores for each word in the query to all the words in the answer. 
%We then select the highest scored pair of query word and answer word and finally add the highest relevance score for all the words in the query. 
%We are referring this method as one-to-one pairing because we just consider the score with the most relevant term in the answer. 
This baseline has been used for other NLP tasks, such as  document matching \cite{Kim2015MindTG} and sentence similarity \cite{kenter2015short}. 

{\flushleft{\bf One-to-all:}} We additionally compare against a model without an alignment threshold, reducing Equation \ref{eqn:align} to:
% \vspace{-4mm}
\begin{small}
\begin{equation}
\mathit{align}(q_i, A) = \sum_{k=1}^{m} \frac{1}{k} \cdot \mathit{cosSim}(q_i, a^+_{q_i,k})
\end{equation}
\end{small}
where $m$ is the number of words in the answer candidate.
%We compute similarity score of each word in the query to all the words in the candidate answer. 
%Hence, we sum the similarity scores of i*n words where i is the total number of query words and n is the total number of candidate answer words.

\subsection{Supervised Model Comparisons:}
For each QA dataset, we compare against previous supervised systems.

{\flushleft{\bf WikiQA:}} For WikiQA, we compare against strong RNN and attention based QA systems. \citet{jurczyk2016selqa} use multiple RNN models with attention pooling. %, achieving maximum of 67.47 percent MAP value. 
 Both \citet{yin2015abcnn} and \citet{dos2016attentive}  use similar approaches of attention layers over CNN's and RNN's.% have been used by  etc for WikiQA achieving atmost 69.21 percent MAP score. 
 ~\citet{miller2016key} use key value memory networks using Wikipedia as the knowledge base and \citet{Tymoshenko2017RankingKF} employed a hybrid of Tree Kernals and CNNs.
%Few other neural architectures like \cite{miller2016key}, \cite{Tymoshenko2017RankingKF} have marginally improved the performance compared to \cite{yin2015abcnn}.   

{\flushleft{\bf YA:}}  For the YA dataset, \citet{jansen14} use discourse, lexical semantic, and IR features in a linear SVM. \citet{fried2015higher} also use a linear SVM but with higher-order alignment features (i.e., ``multi-hop'' alignment). \citet{Bogdanova2016ThisIH} used learned representations of questions and answers in a feed-forward NN and \citet{Liu2017IfYC} use explicit features alongside recurrent NN representations of questions and answers.

{\flushleft{\bf ScienceQA:}}  We compare against the most recently published works for this dataset.  \citet{khot2017tupleinf} employ Integer Linear Programming with a knowledge base of tuples to select the correct answers.  \citet{sharp2017tell} use a combination of learned and explicit features in a shallow NN to simultaneously rerank answers and their justifications.

{\flushleft{\bf ARC:}} Several supervised neural baseline models\footnote{\url{http://data.allenai.org/arc/leaderboard}} have been re-implemented by \citet{clark2018think} for use on the ARC dataset: Decomposed Graph Entailment Model (DGEM)\cite{khot2018scitail}, Bi-directional Attention Flow (BiDAF) \cite{seo2016bidirectional} and Decomposable Attention model \cite{parikh2016decomposable}.

\begin{table}
\begin{center}
\begin{footnotesize}
\begin{tabular}{lcccc|c}
	\hline
	Dataset & $K^+$ & $K^-$ & $\lambda$ & $N$ (Optional) & Q:A  \\
	\hline
    %WikiQA & 5 & 0.4 & Q-3.23,A-13.59 &Q-3.16,A-13.33 \\
    WikiQA & 5 & 1 & 0.4 & N/A & 1:4  \\
% 	\hline
    %ScienceQA & 1 & 0.4 & Q-15.20,A-7.26 & Q-14.56,A-7.19 \\
    ScienceQA & 1 & 1 & 0.4 & 5 (Untuned) & 2:1  \\
% 	\hline
    Yahoo QA & 3 & 0 & -- & N/A & 1:5 \\
% 	\hline
    ARC Challenge & 1 & 0 & -- & 5 (Untuned) & 1:1 \\
    ARC Easy & 1 & 0 & -- & 32 (Tuned) & 1:1 \\
\end{tabular}
\end{footnotesize}
\end{center}
\caption{Tuned values for hyperparameters along with approximate ratios of average number of terms in questions versus answers across the dataset.  $K^+$ and $K^-$ are the number of most and least similar terms, $\lambda$ is the weight of the negative information, and $N$ is the number of IR retrieved justifications (only relevant in the multiple choice datasets, and only tuned in ARC Easy (see Section \ref{sec:tuning}). }
\vspace{-7mm}
\label{tab:tuning}
\end{table}

\subsection{Tuning}
\label{sec:tuning}
As described in Section \ref{sec:approach}, our proposed model has just 3 hyperparameters: $K^+$, the number of positive alignments for each question term; $K^-$, the number of negative alignments; and $\lambda$, the weight assigned to the negative information.  We tuned each of these on development and show the selected values in Table \ref{tab:tuning}.  

We hypothesize that these empirically determined best values for the hyperparameters are correlated with the ratio between the average length of questions and answers across the dataset\footnote{Length statistics were calculated after stop word removal.}  %bs: removed for space
(also shown in Table \ref{tab:tuning}).
That is, in the question sets where answers tend to be several times longer than questions, more alignments per question term were useful. This is in direct contrast with the Science dataset, where questions are typically twice as long as answers.

When running the model on the ARC easy set, we found that commonly the top five justifications retrieved for each of the answer candidates were identical, which prevented our model from differentiating between candidates.  Therefore, \textit{for this dataset only} we also tuned the number of justifications used by the model, i.e., $N$ in Table \ref{tab:tuning}, to 32 (using the training and development set) in order to enable retrieval of distinct justifications.

% \textcolor{red}{Becky, can you help in putting these in nice words, I am unable to think of a good way to present this.}

% \textcolor{blue}{Specifically for ARC easy set, number of justifications in the same setup as \cite{sharp2017tell} affected the overall performance of the model. We had run our experiments till 40 justifications for each candidate answers with best performance on 40th justification. After error analysis, we found the top justifications to be the same for all 4 candidate answers and thus our model's performance increased continuously aggregating scores from more justifications.  }

\begin{table}[t]
\begin{center}
%\begin{footnotesize}
\begin{footnotesize}
\begin{tabular}{lllll}
\hline
% 	&	&	\multicolumn{2}{c|}{Dev} & \multicolumn{2}{c}{Test} \\ 
\# & Supervised & Model &  MAP \\ 
\hline
%& Baselines & \\
%\hline
1	& No &	Wgt Word Cnt \cite{yang2015wikiqa} & 50.99 	\\
2	& Yes	& LCLR \cite{yang2015wikiqa} & 59.93	\\
3	& No	& Our model (One-to-one) 	& 	62.77\\
4	& No	& Our model (One-to-all)  & 	60.91\\
\hline
%& Supervised Models	&	& \\
%\hline
5	& Yes	& \citet{yang2015wikiqa} CNN+Cnt  & 65.20  &\\
6	& Yes	& \citet{jurczyk2016selqa}RNN-1way  &  66.64  & \\ 
7	& Yes	& \citet{jurczyk2016selqa} RNN-Attention\_pool & 67.47 & \\
8	& Yes	& \citet{dos2016attentive} & 68.86	&  \\
9	& Yes	& \citet{yin2015abcnn} &	69.21  &  \\
10	& Yes	& \citet{miller2016key} &	70.69  &  \\
11	& Yes	& \citet{Tymoshenko2017RankingKF}&72.19	&  \\
\hline
12	& No	&	Our final model 	& 64.02$^{*\dagger}$ \\ 
\end{tabular}
\end{footnotesize}
%\vspace{-2mm}
\caption{ \footnotesize Performance on the WikiQA dataset, measured by mean average precision (MAP), for other baselines (both supervised and unsupervised), recent supervised systems, and finally our approach. 
$^*$ and $^\dagger$ indicate that the difference between the model and the One-to-one and One-to-all baselines (respectively) is statistically significant ($p < 0.05$), as determined through a one-tailed bootstrap resampling test with 10,000 iterations.  } 
\label{tab:wikiqa}
\vspace{-5mm}
\end{center}
\end{table}

% \vspace{-10mm}

\begin{table}[t]
\begin{center}
\begin{footnotesize}
\begin{tabular}{llll}
\hline
\# & Supervised & Model & P@1 \\ 
\hline
1	& No &	BM25		   	& 18.60 	\\
2	& No &	CR \cite{jansen14}		   	&  19.57 \\
3	& Yes &	CR + LS \cite{jansen14}		   	&  26.57 \\
4	& No & Our model (One-to-one) 			& 	28.41  \\
5	& No & Our model (One-to-all) 			& 20.17	\\
\hline
6 	& Yes &	\citet{jansen14} 	       & 30.49 \\
7 	& Yes &	\citet{fried2015higher}	 & 33.01 \\ 
8	& Yes &	\citet{Bogdanova2016ThisIH}  		& 37.17 \\
9	& Yes &	\citet{Liu2017IfYC}  		& 38.74 \\
\hline
10	& No &	Our final model 	 &	32.93$^*\dagger$ \\ 
\end{tabular}
\end{footnotesize}
\caption{\footnotesize Performance on the Yahoo! Answers dataset, measured by precision-at-one (P@1).  Significance is indicated as described in Table \ref{tab:wikiqa}.
}
\vspace{-6mm}
\label{tab:yahoo}

\end{center}
\end{table}
% \vspace{-0mm}
\begin{table}[t]
\begin{center}
%\begin{footnotesize}
\begin{footnotesize}
\begin{tabular}{llll}
\hline
% 	&	 & \multicolumn{1}{c}{Test} \\ 
\# & Supervised & Model &  P@1 \\ 
\hline
%& Baselines & \\
%\hline
1	& No &	BM25					& 	39.75\\
2	& No &	Our model (One-to-one) 			& 	46.38\\
3	& No &	Our model (One-to-all) 			& 	34.13\\
\hline
%& Supervised Models	&	& \\
%\hline
4 & Yes &	\citet{khot2017tupleinf} & 46.17 \\ 
5 & Yes & \citet{sharp2017tell} & 53.30 \\
\hline
%& Our Work & &\\
%\hline
6	& No &	Our final model 	 &	47.00$^\dagger$ \\ 
\end{tabular}
\end{footnotesize}
% \vspace{-5mm}
\caption{ \footnotesize Performance on the 8th grade science dataset, measured by precision-at-one (P@1).  Significance is indicated as in Table \ref{tab:wikiqa}.
} 
\label{tab:science}
\vspace{-8mm}
\end{center}
\end{table}

\begin{table}[t]
\begin{center}
%\begin{footnotesize}
\begin{footnotesize}
\begin{tabular}{lllll}
\hline
% 	&	 & \multicolumn{1}{c}{Test} \\ 
 &  &  							&  Easy 	& Challenge \\ 
\#	&	Supervised	&	Model	&  P@1		& P@1	\\
\hline
%& Baselines & \\
%\hline
1	& No &	AI2 IR Solver \cite{clark2018think}, reported	& 59.99\tablefootnote{Performance results on both the Easy and Challenge sets reflect updated numbers from correspondence with the authors.} & 23.98\\
%2	& No &	AI2 IR Solver (\citet{clark2016combining}), \textcolor{red}{our implementation} & 	49.01 & 23.74 \\
2	& No &	Our reimpl. AI2 IR Solver \cite{clark2016combining,clark2018think} \tablefootnote {Following the description provided just in \citet{clark2016combining},  our re-implementation did not include any additional steps mentioned in \citet{clark2018think} which may have improved the performance, such as filtering out overly long justifications or those with negation etc.} & 	49.01 & 23.74 \\
3	& No &	Our model (One-to-one) 			& 	58.36 & 26.56\\

4	& No &	Our model (One-to-all) 			& 	48.04 & 25.45 \\

\hline

5 & Yes & DA (for ARC) \cite{parikh2016decomposable,clark2018think} & 58.27 & 24.34  \\

6 & Yes & BiDAF (for ARC) \cite{seo2016bidirectional,clark2018think}& 50.11 & 26.54  \\ 

7 & Yes & DGEM (for ARC) \cite{khot2018scitail,clark2018think}& 58.97 & 27.11 \\ 

8 & Yes & DGEM-OpenIE (for ARC) \cite{khot2018scitail,clark2018think}& 57.45 & 26.41 \\

\hline

9 & No & Our final model & 58.36 & 26.56\\

\end{tabular}
\end{footnotesize}
% \vspace{-5mm}
\caption{ \footnotesize Performance on the ARC dataset, measured by precision-at-one (P@1).} 
\label{tab:ARC}
\vspace{-8mm}
\end{center}
\end{table}

\section{Results and Discussion}
\label{sec:results}

% Main results -- compare to baselines
We evaluate our approach on four distinct QA datasets: WikiQA, Yahoo! Answers (YA), 8th grade science dataset (ScienceQA) and ARC dataset.  These results are shown in Tables \ref{tab:wikiqa}, \ref{tab:yahoo}, \ref{tab:science} and \ref{tab:ARC}.
In Science and Yahoo! QA, our approach significantly outperforms BM25 ($p < 0.05$)\footnote{All statistical significance determined through one-tailed bootstrap resampling with 10,000 iterations.}, demonstrating that incorporating lexical semantic alignments between question terms and answer terms (i.e., going beyond strict lexical overlap) is beneficial for QA. LCLR \cite{yih13,yang2015wikiqa} is considered to be a stronger baseline for WikiQA, and our model outperforms it by +4.10\% MAP. 

Further, we compare our full model %(which includes both a one-to-many alignment as well as negative information) %bs: removed for space
with both a single alignment approach (i.e., one-to-one) as well as a maximal alignment (i.e., one-to-all) approach. % full vs 1-to-1 
In all datasets except for the ARC datasets, our full model performed better than the single alignment approach; in both WikiQA and YA this difference was significant ($p < 0.05$). 
% full vs 1-to-all 
In case of the ARC dataset, our final model was identical to a single alignment model since multi-alignment and negative information did not improving performance development.

Our full model was also significantly better than the one-to-all baseline in all models ($p < 0.05$).
This demonstrates that including additional context in the form of multiple alignments is useful, but that there is a ``Goldilocks'' zone, and going beyond that is detrimental. %\todo{mention length effect here} \todo{talk about yahoo?} % ms: no space... discuss in camera ready
We note that while the negative alignment boosted performance, in none of the datasets was its contribution significant individually.

% \textcolor{red}{Becky: Vikas I would remove this paragraph.  If we go the route of describing it as an additional hyperparameter in only the ARC easy, then we'll talk about it just in the tuning section.  And I guess we'd say 32 instead of 40... }
% \textcolor{blue}{The number of justifications did not vary the performance of 8th grade scienceQA on dev set as opposed ARC easy dataset. Table \ref{tab:ARC} shows that performance improvement of our proposed model (utilizing 40 justifications) over strong baseline of BM25 model with \citet{clark2016combining} settings of IR solver model.}

\begin{figure}[t]
\begin{center}
\includegraphics[width=0.41\textwidth]{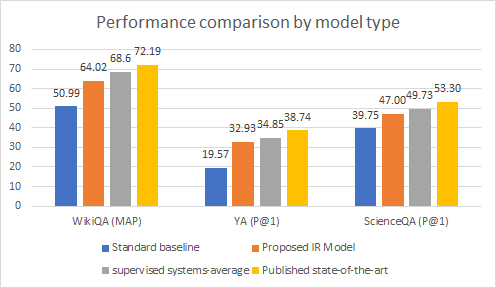}
\vspace{-3mm}
\caption{\footnotesize Histogram depicting the performance across three QA datasets of the standard baselines (shown in blue), our proposed unsupervised model (orange), the average of recently proposed supervised systems (grey), and the current state of the art (yellow).  Notably, our model exceeds the standard baselines and approaches the mean of the supervised systems in all three datasets.
%gap between our model (orange) and the standard baselines and closeness to mean of supervised models and state of the art performance in all 3 datasets.
}
\vspace{-5mm}
\label{fig:Graph}
\end{center}
\end{figure}

%\citet{wang2016sentence} presented a supervised model using CNN to extract features from both negative and positive information after computing semantic similarity between two sentences similar to our proposed approach. Since we have used an unsupervised and only 1 dimension weight Lambda (contrary to n dimension weights used by CNN cell) to weight negative information, we achieve significant improvement on validation data because of tuning but not on test data. On contrary, wang reported significant improvement with negative information. 

Perhaps more interestingly, despite its simplicity, lack of parameters, and unsupervised nature, our approach either beats or approaches many much more complex supervised systems with steep training costs (e.g., attention-based RNNs), showing we can come closer to bridging the performance gap between simple baselines and complex systems using straightforward approaches, as illustrated in Figure \ref{fig:Graph}.
We suspect that our proposed approach would also be complementary to several of the more complex systems (particularly those without IR components \cite[e.g.][]{khot2017tupleinf}), which would allow for additional gains through ensembling.

\section{Conclusion}
\label{sec:conclusion}

We introduced a fast and simple, yet strong, unsupervised baseline approach for QA that uses pre-trained word embeddings to produce one-to-many alignments between question and answer words, capturing both positive and negative alignments. Despite its simplicity, our approach considerably outperforms all current baselines, as well as several complex, supervised systems, approaching state-of-the-art performance on three QA tasks. 
Our work suggests that simple alignment strategies remain strong contenders for QA, and that the QA community would benefit from such stronger baselines for more rigorous analyses. 
%BS - addressed: [RESPONSE TODO: In conclusion , emph unsupervised, fast]